\documentclass[
aps,%
12pt,%
final,%
notitlepage,%
oneside,%
onecolumn,%
nobibnotes,%
nofootinbib,
superscriptaddress,%
noshowpacs,%
centertags]%
{revtex4}
\RequirePackage{graphicx}

\def\refitem#1{\relax}

\begin{document}
\title{Measuring Dynamical K/$\pi$ and p/$\pi$ Fluctuations in Au+Au Collisions from the STAR Experiment}

\author{\firstname{Terence} \surname{Tarnowsky}  (for the STAR Collaboration)}
\email{tarnowsk@nscl.msu.edu}
\affiliation{National Superconducting Cyclotron Laboratory, Michigan State University,
East Lansing, MI 48824, USA}

\begin{abstract}
Results from new measurements of dynamical $K/\pi$ and $p/\pi$ ratio fluctuations are presented. Dynamical fluctuations in global conserved quantities such as baryon number, strangeness, or charge may be observed near a QCD critical point. The STAR experiment has previously acquired data in Au+Au collisions at the energies $\sqrt{s_{NN}}$ = 200, 130, 62.4, and 19.6 GeV and Cu+Cu collisions at $\sqrt{s_{NN}}$ = 200, 62.4, and 22.4 GeV. The commencing of a QCD critical point search at RHIC has extended the reach of possible measurements of dynamical $K/\pi$ and $p/\pi$ ratio fluctuations from Au+Au collisions to lower energies. New results are compared to previous measurements and to theoretical predictions from the UrQMD model.
\end{abstract}

\maketitle

\section{Introduction}

Fluctuations and correlations are well known signatures of phase transitions. In particular, the quark/gluon to hadronic phase transition may lead to significant fluctuations \cite{Koch1}. In 2010, the Relativistic Heavy Ion Collider (RHIC) began a program to search for the QCD critical point. This involves an ``energy scan'' of Au+Au collisions from top collision energy ($\sqrt{s_{NN}}$ = 200 GeV) down to energies as low as $\sqrt{s_{NN}}$ = 7.7 GeV \cite{STARBES}. This critical point search will make use of the study of correlations and fluctuations, particularly those that could be enhanced during a phase transition that passes close to the critical point \cite{kurtosis}. Particle ratio fluctuations are an observable that has already been studied as a function of energy and system-size. Combining these previous measurements with new results from the energy scan will provide additional information to assist in locating the QCD critical point.

Dynamic particle ratio fluctuations, specifically fluctuations in the $K/\pi$ and $p/\pi$ ratio, can provide information on the quark-gluon to hadron phase transition \cite{Strangeness1, Strangeness2, Strangeness3}. The variable used to measure these fluctuations is referred to as $\nu_{dyn}$. $\nu_{dyn}$ was originally introduced to study net charge fluctuations \cite{nudyn1, nudyn2}. $\nu_{dyn}$ quantifies deviations in the particle ratios from those expected for an ideal statistical Poissonian distribution. The definition of $\nu_{dyn,K/\pi}$ (describing fluctuations in the $K/\pi$ ratio) is,
\begin{eqnarray}
\nu_{dyn,K/\pi} = \frac{<N_{K}(N_{K}-1)>}{<N_{K}>^{2}}
+ \frac{<N_{\pi}(N_{\pi}-1)>}{<N_{\pi}>^{2}}
- 2\frac{<N_{K}N_{\pi}>}{<N_{K}><N_{\pi}>}
\label{nudyn}
\end{eqnarray}
where $N_{K}$ and $N_{\pi}$ are the number of kaons and pions in a particular event, respectively. In this proceeding, $N_{K}$ and $N_{\pi}$ are the total charged multiplicity for each particle species. A formula similar to (\ref{nudyn}) can be constructed for $p/\pi$ and other particle ratio fluctuations. By definition, $\nu_{dyn}$ = 0 for the case of a Poisson distribution of kaons and pions. It is also largely independent of detector acceptance and efficiency in the region of phase space being considered \cite{nudyn2}. An in-depth study of $K/\pi$ fluctuations in Au+Au collisions at $\sqrt{s_{NN}}$ = 200 and 62.4 GeV was previously carried out by the STAR experiment \cite{starkpiprl}.

Earlier measurements of particle ratio fluctuations utilized the variable $\sigma_{dyn}$ \cite{NA49}, 
\begin{equation}
\sigma_{dyn} = sgn(\sigma_{data}^{2}-\sigma_{mixed}^{2})\sqrt{|\sigma_{data}^{2}-\sigma_{mixed}^{2}|}
\label{signudyn}
\end{equation}
where $\sigma$ is the relative width of the $K/\pi$ or $p/\pi$ distribution in either real data or mixed events. It has been shown that $\nu_{dyn}$ is a first order expansion of $\sigma_{dyn}$ about the mean of its denominator \cite{jeon,baym,sdasthesis}. The two variables are related as $\sigma_{dyn}^{2} \approx \nu_{dyn}$.

\section{Experimental Analysis}

The data presented here for $K/\pi$ and $p/\pi$ fluctuations was acquired by the STAR experiment at RHIC from minimum bias (MB) Au+Au collisions at center-of-mass collision energies ($\sqrt{s_{NN}}$) of 200, 130, 62.4, 39, 19.6, and 7.7 GeV \cite{STAR}. The main particle tracking detector at STAR is the Time Projection Chamber (TPC) \cite{STARTPC}. All detected charged particles in the pseudorapidity interval $|\eta| < 1.0$ were measured. The transverse momentum ($p_{T}$) range for pions and kaons was $0.2 < p_{T} < 0.6$ GeV/$c$, and for protons was $0.4 < p_{T} < 1.0$ GeV/$c$. Charged particle identification involved measured ionization energy loss ($dE/dx$) in the TPC gas and total momentum ($p$) of the track. The energy loss of the identified particle was required to be less than two standard deviations (2$\sigma$) from the predicted energy loss of that particle at a particular momentum. A second requirement was that the measured energy loss of a pion/kaon was more than 2$\sigma$ from the energy loss prediction of a kaon/pion, and similarly for $p/\pi$ measurements. The centralities used in this analysis account for 0-10, 10-20, 20-30, 30-40, 40-50, 50-60, 60-70, and 70-80\% of the total hadronic cross section.

Data was also analyzed using the recently completed Time of Flight (TOF) detector \cite{STARTOF}. The TOF is a multi-gap resistive plate chamber (MRPC) detector. Particle identification was carried out using the time-of-flight information for a track along with its momentum, determined by the TPC. From these two quantities the particle's mass can be calculated. Mass-squared ($m^{2}$) cuts used for particle ID are: $0.001 \leq m^{2} \leq 0.07$ for pions, $0.21 \leq m^{2} \leq 0.29$ for kaons, and $0.8 \leq m^{2} \leq 1.0$ for protons. In this proceeding, the $p_{T}$ range presented for TOF results is the same as that from the TPC, $0.2 < p_{T} < 0.6$ GeV/$c$. Ultimately, use of the TOF will provide access to particle identification out to $\approx$ 1.6 GeV/$c$ for pions and kaons.

\section{Results and Discussion}

The STAR experiment has recently acquired data at three new energies. The first phase of the QCD critical point search involved Au+Au collisions at $\sqrt{s_{NN}}$ = 39, 11.5, and 7.7 GeV. The next phase will include Au+Au collisions at $\sqrt{s_{NN}}$ = 27 and 18 GeV. As of August 2010, STAR has newly measured dynamical $K/\pi$ fluctuations at $\sqrt{s_{NN}}$ = 39 GeV and $p/\pi$ fluctuations at both $\sqrt{s_{NN}}$ = 39 and 7.7 GeV. 

\begin{figure}[h]
\centering
\includegraphics[width=1.000\textwidth]{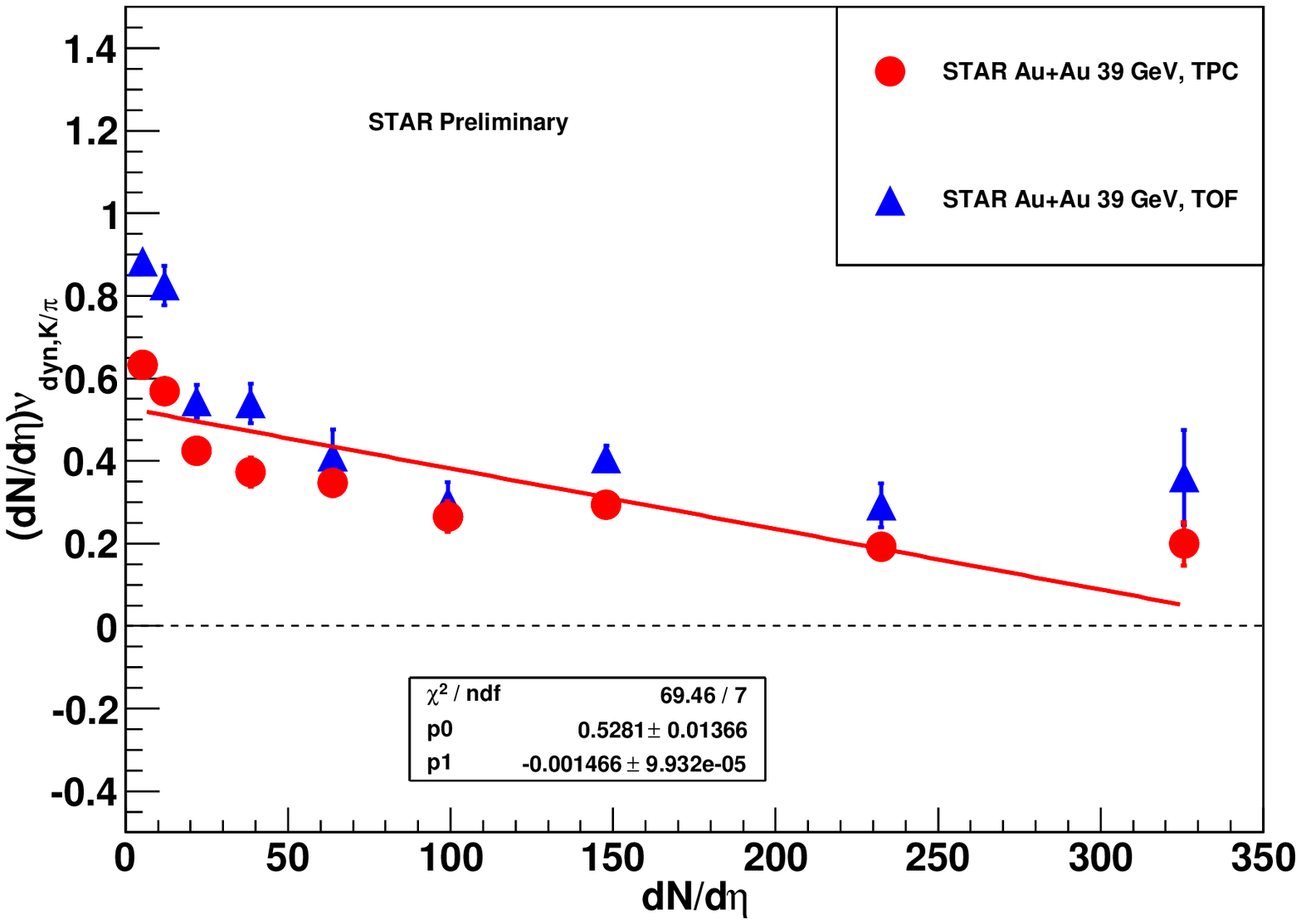} 
\caption{Results for the measurement of $\nu_{dyn}$ for $K/\pi$ fluctuations as measured by the STAR TPC (red circles) and TOF (blue circles) from Au+Au collisions at $\sqrt{s_{NN}}$ = 39 GeV, scaled by uncorrected charged particle multiplicity, $dN/d\eta$. A linear fit to the TPC data and its parameters are also shown. All errors are statistical.}
\label{nudynkpi39}
\end{figure}

Figure \ref{nudynkpi39} shows the first measurement of $\nu_{dyn}$ for $K/\pi$ fluctuations as a function of centrality, scaled by the charged particle multiplicity, $dN/d\eta$, from Au+Au collisions at $\sqrt{s_{NN}}$ = 39 GeV. The measured $dN/d\eta$ is a raw value, not corrected for efficiency and acceptance. Scaling by $dN/d\eta$ removes the 1/$N_{ch}$ dependence of $\nu_{dyn}$ \cite{nudyn1}, where $N_{ch}$ is the charged particle multiplicity. There are two sets of data points. For the red circles, only particles that passed a TPC dE/dx particle identification cut were used. For the blue circles, only TOF mass-squared particle identification was used. The errors shown are statistical. Dynamical $K/\pi$ fluctuations are always positive. 
Accounting for track matching efficiency and detector acceptance, the TOF detection efficiency is approximately $\frac{2}{3}$ that of the TPC. Differences between the two data samples are still under investigation.

$\nu_{dyn,K/\pi}$ was found to scale with $dN/d\eta$ in peripheral and semi-central Au+Au collisions at $\sqrt{s_{NN}}$ = 200 and 62.4 GeV \cite{starkpiprl}. The TPC data in Figure \ref{nudynkpi39} was fit with a linear function and the resulting slope is (-1.47 $\pm$ 0.01) $\times$ $10^{-3}$. The linear scaling of $\nu_{dyn,K/\pi}$ may not hold in very peripheral collisions at $\sqrt{s_{NN}}$ = 39 GeV.

\begin{figure}[h]
\centering
\includegraphics[width=1.000\textwidth]{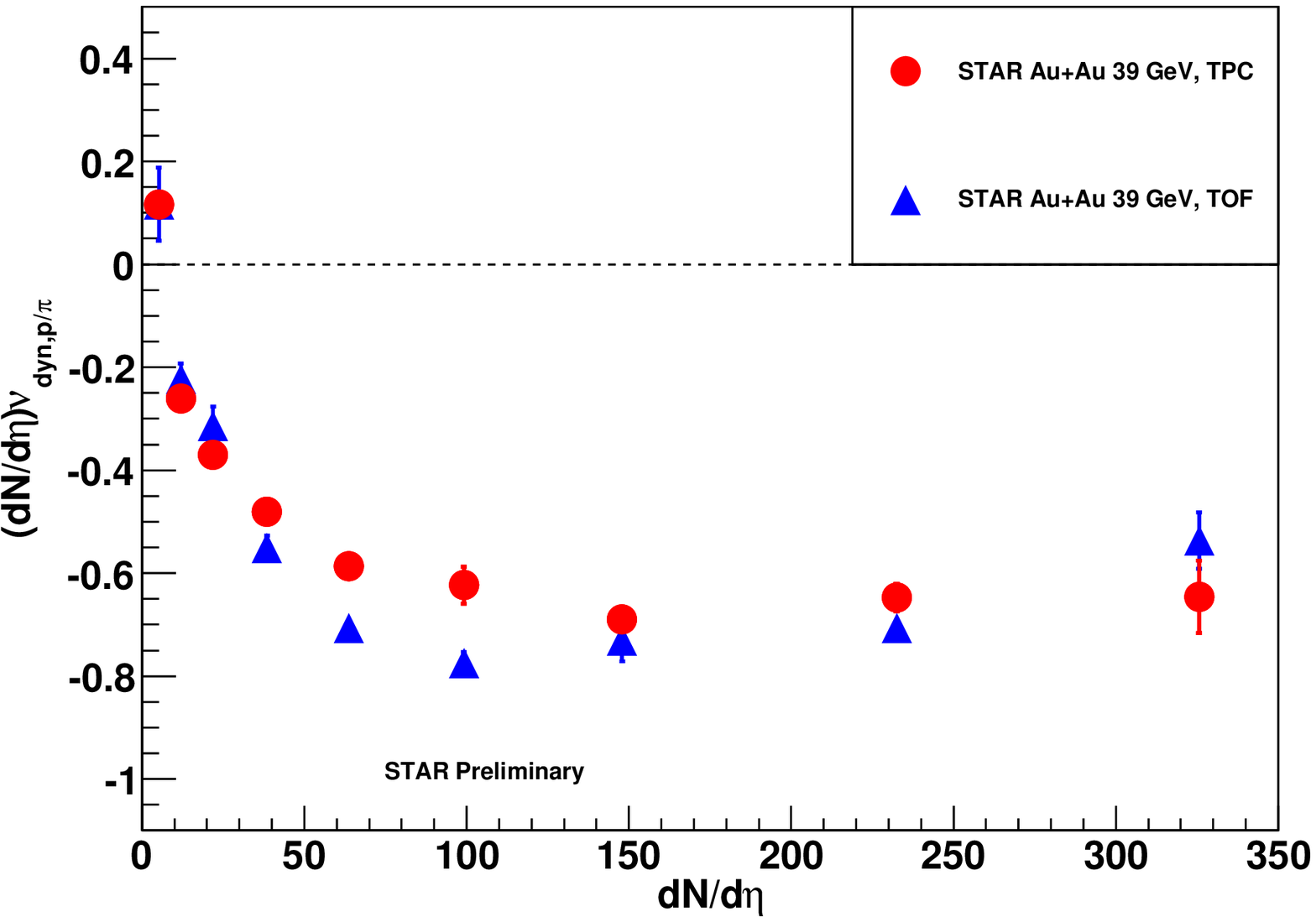} 
\caption{Results for the measurement of $\nu_{dyn}$ for $p/\pi$ fluctuations as measured by the STAR TPC (red circles) and TOF (blue circles) from Au+Au collisions at $\sqrt{s_{NN}}$ = 39 GeV, scaled by uncorrected charged particle multiplicity, $dN/d\eta$. All errors are statistical.}
\label{nudynppi39}
\end{figure}

The first measurement of $\nu_{dyn}$ for $p/\pi$ fluctuations in Au+Au collisions at $\sqrt{s_{NN}}$ = 39 GeV is shown in Figure \ref{nudynppi39}. As in Figure \ref{nudynkpi39}, they are presented as a function of centrality, scaled by the raw charged particle multiplicity, $dN/d\eta$. The red circles are particles that passed a TPC dE/dx cut, while the blue circles are particles that passed a TOF mass-squared cut. Unlike $\nu_{dyn,K/\pi}$, $\nu_{dyn,p/\pi}$ is always negative as a function of centrality. This implies that correlated production of protons and pions, dominates.

\begin{figure}[h]
\centering
\includegraphics[width=1.000\textwidth]{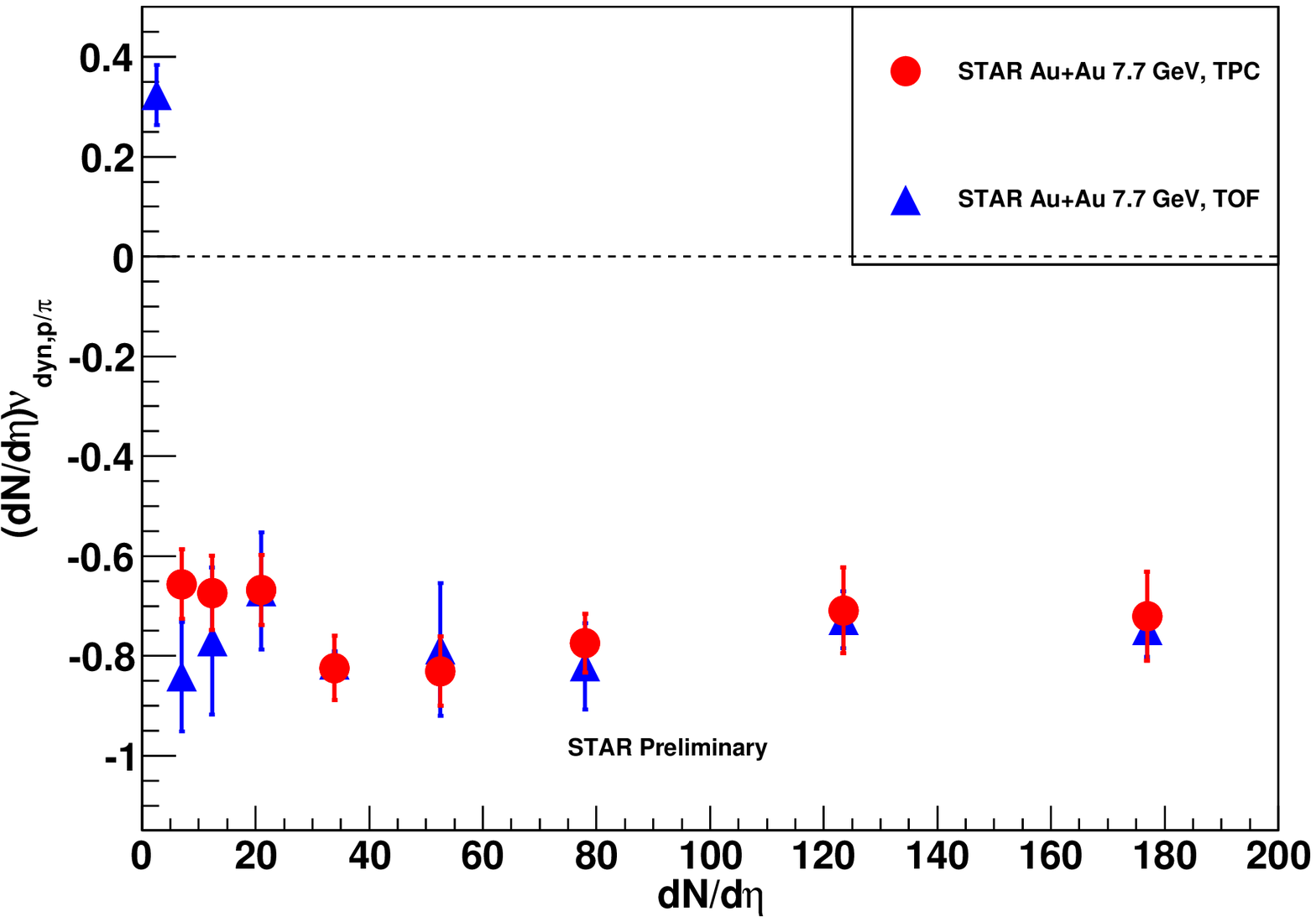} 
\caption{Results for the measurement of $\nu_{dyn}$ for $p/\pi$ fluctuations as measured by the STAR TPC (red circles) and TOF (blue circles) from Au+Au collisions at $\sqrt{s_{NN}}$ = 7.7 GeV, scaled by uncorrected charged particle multiplicity, $dN/d\eta$. All errors are statistical.}
\label{nudynppi7}
\end{figure}

Figure \ref{nudynppi7} is the first measurement of dynamical $p/\pi$ fluctuations in Au+Au collisions at $\sqrt{s_{NN}}$ = 7.7 GeV. This is currently the lowest energy data available at RHIC. As with dynamical fluctuations at $\sqrt{s_{NN}}$ = 39 GeV, $\nu_{dyn}$ is plotted as a function of centrality and scaled by the uncorrected charged particle multiplicity, $dN/d\eta$. At this low energy, the antibaryon/baryon ratio is $\sim$ 0.01 \cite{NA49_pbarp}. Thus, the measurement of $p/\pi$ fluctuations at this energy is dominated by protons. Measurements using both TPC particle identification (red circles) and TOF particle identification (blue circles) are strongly negative and approximately independent of centrality. This can be compared to Figure \ref{nudynppi39}, where $\nu_{dyn,p/\pi}$ exhibits a centrality dependence in peripheral collisions.

\begin{figure}[h]
\centering
\includegraphics[width=1.000\textwidth]{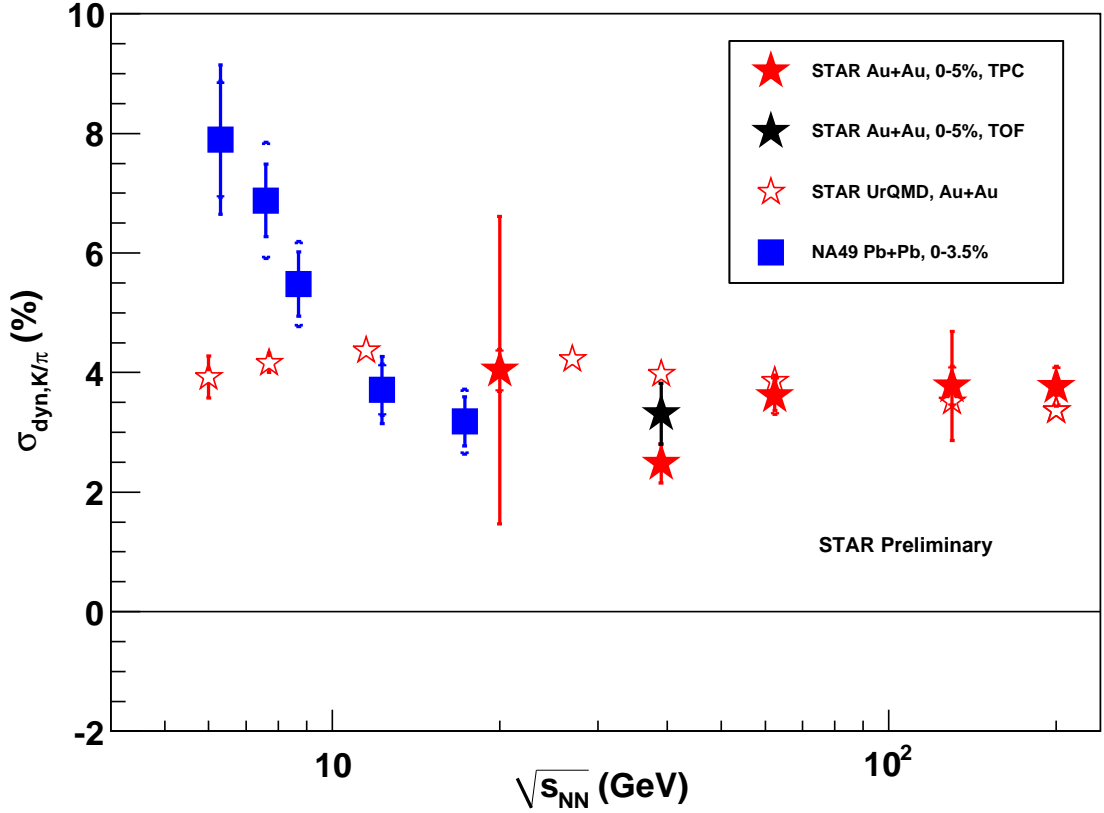} 
\caption{$K/\pi$ fluctuations expressed as $\sigma_{dyn}$, as a function of incident energy. Data is shown from both the STAR (solid red stars) and NA49 (solid blue squares) experiments from central collisions, Au+Au, 0-5\% for STAR and Pb+Pb, 0-3.5\% for NA49. Also shown are model calculations from UrQMD using the STAR (open red stars) experimental acceptance. Errors include both statistical and systematic effects, except for the STAR points at $\sqrt{s_{NN}}$ = 39 GeV, which have only statistical errors.}
\label{kpi_excitation}
\end{figure}

The energy dependence of dynamical $K/\pi$ fluctuations is summarized in Figure \ref{kpi_excitation}. The calculated dynamical $K/\pi$ fluctuation results from STAR are converted to $\sigma_{dyn}$ using the relationship $\sigma_{dyn}^{2} \approx \nu_{dyn}$. Errors were also propagated from $\nu_{dyn}$ to $\sigma_{dyn}$. There is a strong decrease with increasing incident energy for the NA49, Pb+Pb 0-3.5\% results (solid blue squares). The STAR results (solid red stars) for 0-5\% Au+Au at $\sqrt{s_{NN}}$ = 19.6, 62.4, 130, and 200 GeV are independent of energy within errors. The newly measured $K/\pi$ fluctuations from Au+Au collisions at $\sqrt{s_{NN}}$ = 39 GeV are also included and are similar in magnitude to the other STAR results.
Also shown is a prediction from the UrQMD model using the STAR detector acceptance (open red stars). The UrQMD model predicts a value of $\sigma_{dyn,K/\pi}$ at approximately 4\% for all energies. 

\begin{figure}[h]
\centering
\includegraphics[width=1.000\textwidth]{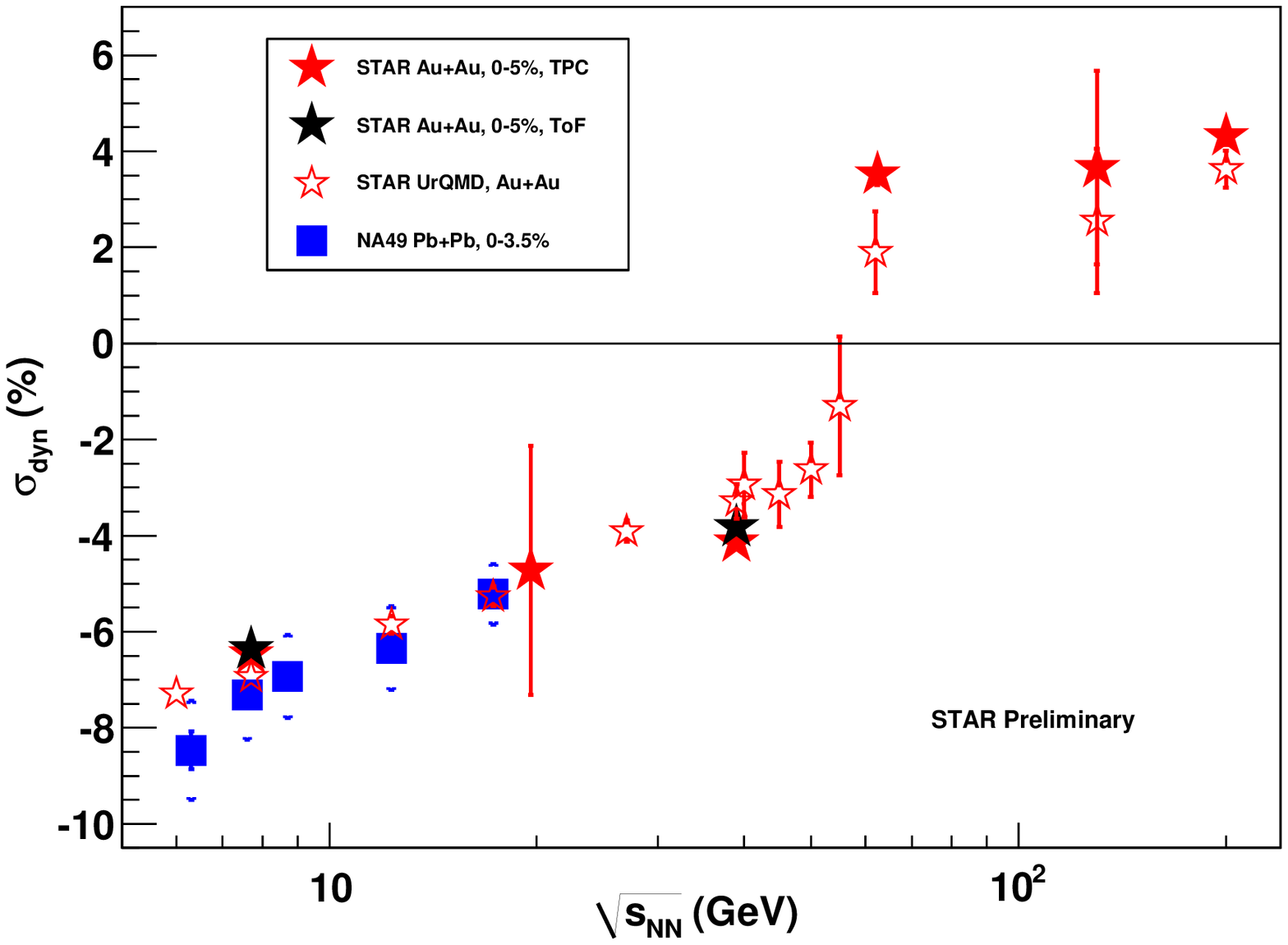} 
\caption{$p/\pi$ fluctuations expressed as $\sigma_{dyn}$, as a function of incident energy. Data is shown from both the STAR (solid red stars) and NA49 (solid blue squares) experiments from central collisions, Au+Au, 0-5\% for STAR and Pb+Pb, 0-3.5\% for NA49. Also shown are model calculations from UrQMD using the STAR (open red stars) experimental acceptance. The errors bars from the STAR data only include statistical errors.}
\label{ppi_excitation}
\end{figure}

Finally, Figure \ref{ppi_excitation} depicts the energy dependence of dynamical $p/\pi$ fluctuations. As in Figure \ref{kpi_excitation}, $\nu_{dyn,p/\pi}$ has been converted to $\sigma_{dyn}$ using $\sigma_{dyn}^{2} \approx \nu_{dyn}$. The trend exhibited by the data is a positive increase in fluctuations with increasing incident energy for both NA49, Pb+Pb 0-3.5\% results (solid blue squares) and STAR results (solid red stars) for 0-5\% Au+Au collisions. UrQMD model predictions with the STAR experimental acceptance (open red stars) reproduce most of the data across the range of experimental energies. The two new data points from STAR for Au+Au collisions at $\sqrt{s_{NN}}$ = 39 and 7.7 GeV are also shown. The dynamical $p/\pi$ fluctuations from STAR at 7.7 GeV are in close agreement with the NA49 result at $\sqrt{s_{NN}}$ = 7.6 GeV and are self-consistent between the STAR TPC and TOF particle identification in the same $p_{T}$ range. 

\section{Summary}

The first results on particle ratio ($K/\pi$ and $p/\pi$) fluctuations from Au+Au collisions at $\sqrt{s_{NN}}$ = 39 and 7.7 GeV have been presented. The STAR experiment has demonstrated dynamical $K/\pi$ fluctuations that are approximately flat above $\sqrt{s_{NN}}$ = 19.6 GeV and $p/\pi$ fluctuations that increase with energy, starting from $\sqrt{s_{NN}}$ = 7.7 GeV. More precise measurements will be forthcoming with additional experimental data. 
These are the first measurements of dynamical particle ratio fluctuations in Au+Au collisions below injection energy at RHIC. Additional energies will be analyzed as part of the QCD critical point search at RHIC, where STAR will be able to precisely measure dynamical $K/\pi$ and $p/\pi$ fluctuations, providing a comprehensive picture of these fluctuations from $\sqrt{s_{NN}}$ = 7.7-200 GeV.


\newpage

\newpage

\begin{center}
FIGURE CAPTIONS
\end{center}
\begin{enumerate}
\item
Results for the measurement of $\nu_{dyn}$ for $K/\pi$ fluctuations as measured by the STAR TPC (red circles) and TOF (blue circles) from Au+Au collisions at $\sqrt{s_{NN}}$ = 39 GeV, scaled by uncorrected charged particle multiplicity, $dN/d\eta$. A linear fit to the TPC data and its parameters are also shown. All errors are statistical.
\item
Results for the measurement of $\nu_{dyn}$ for $p/\pi$ fluctuations as measured by the STAR TPC (red circles) and TOF (blue circles) from Au+Au collisions at $\sqrt{s_{NN}}$ = 39 GeV, scaled by uncorrected charged particle multiplicity, $dN/d\eta$. All errors are statistical.
\item
Results for the measurement of $\nu_{dyn}$ for $p/\pi$ fluctuations as measured by the STAR TPC (red circles) and TOF (blue circles) from Au+Au collisions at $\sqrt{s_{NN}}$ = 7.7 GeV, scaled by uncorrected charged particle multiplicity, $dN/d\eta$. All errors are statistical.
\item
$K/\pi$ fluctuations expressed as $\sigma_{dyn}$, as a function of incident energy. Data is shown from both the STAR (solid red stars) and NA49 (solid blue squares) experiments from central collisions, Au+Au, 0-5\% for STAR and Pb+Pb, 0-3.5\% for NA49. Also shown are model calculations from UrQMD using the STAR (open red stars) experimental acceptance. Errors include both statistical and systematic effects, except for the STAR points at $\sqrt{s_{NN}}$ = 39 GeV, which have only statistical errors.
\item
$p/\pi$ fluctuations expressed as $\sigma_{dyn}$, as a function of incident energy. Data is shown from both the STAR (solid red stars) and NA49 (solid blue squares) experiments from central collisions, Au+Au, 0-5\% for STAR and Pb+Pb, 0-3.5\% for NA49. Also shown are model calculations from UrQMD using the STAR (open red stars) experimental acceptance. The errors bars from the STAR data only include statistical errors.

\end{enumerate}

\end{document}